# Stokes mode Raman random lasing in a fully biocompatible medium

Venkata Siva Gummaluri, S.R.Krishnan and C.Vijayan*

Department of Physics, Indian Institute of Technology Madras, Chennai, India 600036

*Corresponding Author: cvijayan@iitm.ac.in

*We demonstrate for the first time, Raman random lasing in a continuous-wave (CW) excited, completely biocompatible and biodegradable carrot medium. The Stokes Raman mode of carotene renders carrot as a gain medium whereas random scattering in the natural fibrous cellulose of the carrot structure facilitates random lasing. CW laser induced photoluminescence(PL) threshold and linewidth analysis at the Stokes modes of carotene show a characteristic lasing action with threshold of 130 W/cm$^2$ and linewidth-narrowing with mode Q factors (~1300). Polarization study of output modes reveals that lasing mode mostly retains the source polarization state. A clear and interesting linear temperature dependence of emission intensity is also demonstrated. Easy availability, bio-compatibility, excitation dependent emission wavelength selectivity and temperature sensitivity are hallmarks of this elegant Raman laser medium with a strong potential as an optical source for applications in bio-sensing, imaging and spectroscopy.*

Light sources employing green photonic technologies based on organic media are eco-friendly, bio-degradable, bio–compatible with reduced costs of scalable implementation[1,2]. Random lasers (RL) traditionally use plasmonic/dielectric scatterers, and gain media such as chemical dyes and fluorescent polymers which are generally toxic and non-biodegradable and require optical pumping with high energy/power density for lasing[3–6]. Random lasers based on cellulose fibrous networks have suggested the possibility of an organic/eco-friendly bio-compatible scattering medium[7]. However, the gain media used widely are dyes like rhodamine -6G and -B which are toxic and require careful handling. Continuous-wave (CW) laser pumping could provide an economical approach for random lasers[8]. The idea of Raman random lasing, enhancement of Raman modes with elastic scattering, proposed and implemented by Hokr et al. is worth mentioning here[9,10]. Stable, low threshold Raman mode random lasing based on a ZnS/beta-carotene random medium has been demonstrated by our group[11]. CW laser pumping and the use of biocompatible and eco-friendly materials are of great importance to this field of study. Green materials like carbon dots, which show interesting tunable photoluminescence properties, have been shown to have potential applications as laser materials when incorporated with plasmonic particles and organic dyes[12–14]. Organic bio-pigments like carotenoids and porphyrins are interesting optically active media in view of their visible light absorption property. Raman spectroscopy and other optical properties of carotenoids including ultrafast time resolved studies have been comprehensively explored over the past few decades[15–17]. Although the fluorescence quantum yield of carotenoids is much less compared to standard organic laser dyes, the vibrational spectra can be obtained even with concentrations down to $10^{-8}$ M of these carotenoids, given the excitation line falls within the electronic absorption band[16].

Here we report for the first time, to the best of our knowledge, CW Raman random lasing based on Stokes transitions of carotene in fresh carrot, which is fully biodegradable, eco-friendly and economic. Carrots are rich in Raman active carotene and cellulose scattering fibers, facilitating multiple scattering of photons and contributing to optical amplification for Raman random lasing. Lasing features, Stokes mode polarization state and a linear emission intensity dependence on temperature are discussed.

Lasing and enhancement of Stokes emission in stimulated Raman Scattering media can be characterized by Raman gain of medium. For steady state Raman spectroscopy, as in the present case, gain is determined using the Raman Scattering Cross-section (RSC) and is given by[18],

$$g_R = \frac{4\pi c N(\frac{d\sigma}{d\Omega})}{\hbar\omega_s^3 n_s n_L \delta\nu} \qquad (1)$$

where $g_R$, $c$, $N$, $d\sigma/d\Omega$ are the Raman gain, concentration, number density and RSC of carotene solution respectively. $\omega_s$ and $\delta\nu$ are the Stokes line frequency and line width of Raman mode, and $n_s$ and $n_L$ are the refractive indices of carotene at Stokes and laser-excitation frequencies respectively. For beta-carotene solution of very low concentration, $g_R$ is reported[18] to be $7*10^{-13}$ m/W. This value is similar to $CCl_4$ solution of 1000 times more concentration, rendering carotene to be a very good Raman gain material for exploring Raman random lasing-related mechanism.

The experimental schematic is shown in fig.1(a). Raman spectroscopy and PL studies are carried out on cut carrot pieces using Horiba Jobin-Yvon HR800UV Labram spectrometer with 600 lines/mm grating ($\Delta\lambda$ = 0.05 nm), Ar-ion CW laser (Spectra-Physics, 488 nm, 5 µW - 10 mW), He-Ne laser (633 nm, 1 mW), 50X objective lens (0.5 N.A and 30 µm spot diameter for 488 nm) and thermo-electric cooled CCD high-grade spectroscopy detector. PL spectra are recorded at room temperature with an acquisition time of 1 second. Temperature dependent studies are done using Linkam setup (stage and controller) with controlled pumping of liquid nitrogen. Scanning Electron Microscopy (SEM) image of the cellulose fibers (FEI Quanta 200) in raw carrot is shown in fig.1(b). For preliminary optical characterization, carotene is extracted from carrots under stable heating in ethanol at 45°C for 2 hours. Optical absorption (Jasco V570 UV-vis-NIR) of carotene (Fig. 1(c)) spans from 400 nm to 500 nm which can be used efficiently to excite the molecule. PL spectra (Jasco FP6600 spectrofluorometer) for four standard excitations, 405, 445, 488 and 632 nm, are shown (fig.1(d)) with 405 nm excitation being most efficient.

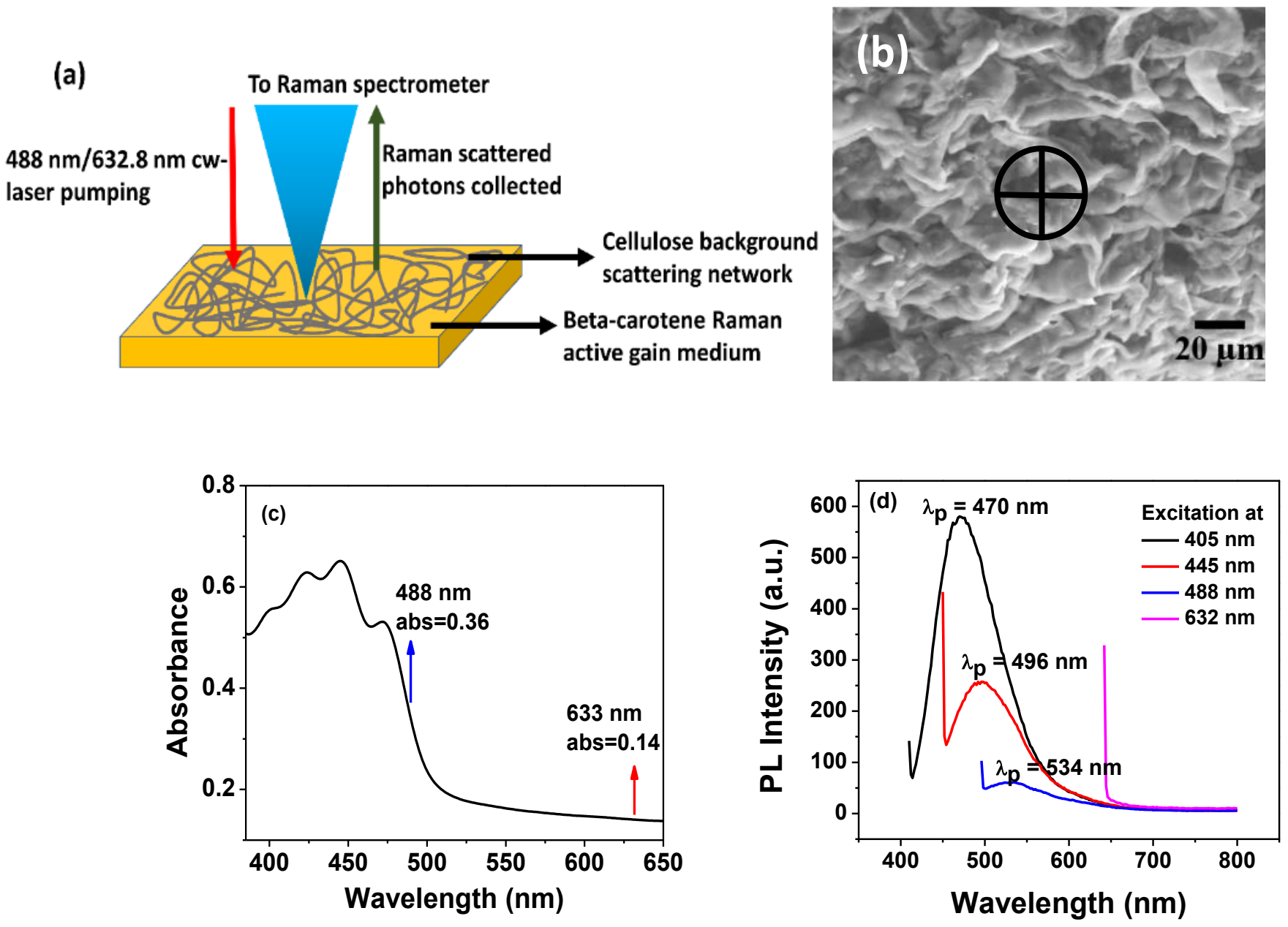

**Figure 1.** (a) Experimental schematic, (b) SEM image of cellulose structures in carrot; the spot of diameter of 488 nm Ar-ion pump laser is shown by the circle inside, (c) optical absorption spectrum of carotene solution,(d) PL spectrum of the solution at standard excitation wavelengths.

We employed coherent backscattering (CBS) technique to characterize the random medium, where the sample is rotated to average the contribution from backscattered intensity. The mean free path $l_t$ is obtained to be 8.8 µm, which signifies diffusive weak localization regime of scattering ($2\pi l_t \gg \lambda$). Thus the cellulose medium provides scattering feedback for amplification of photons at the PL modes of carotene. Hokr and Yakovlev reported that Raman signals can be greatly enhanced by elastic scattering through a turbid or random medium[19], to the amplification due to Raman gain along the increased optical path length.

The Raman mode spectrum, shown in fig.2(a) is in agreement with the reported data for carotene molecule[20]. Three stable and narrow modes are seen at 1000, 1150 and 1525 $cm^{-1}$ corresponding to specific carbon bond vibrations[21]. The three corresponding modes in PL spectrum, shown in fig.2(b), are highly wavelength stable with laser excitation power. The obtained linewidths for 513 nm, 517 nm and 527 nm modes are 0.4 nm, 0.45 nm and 0.5 nm respectively giving a quality factor $Q$ of 1283, 1149 and 1054 ($Q=\lambda_0/\Delta\lambda$, where $\lambda_0$ is the mode peak wavelength and $\Delta\lambda$ is the mode linewidth). We used resonance bandwidth definition of mode $Q$-factor which depends on mode peak wavelength and linewidth[22], and signifies the amount of cavity induced damping.

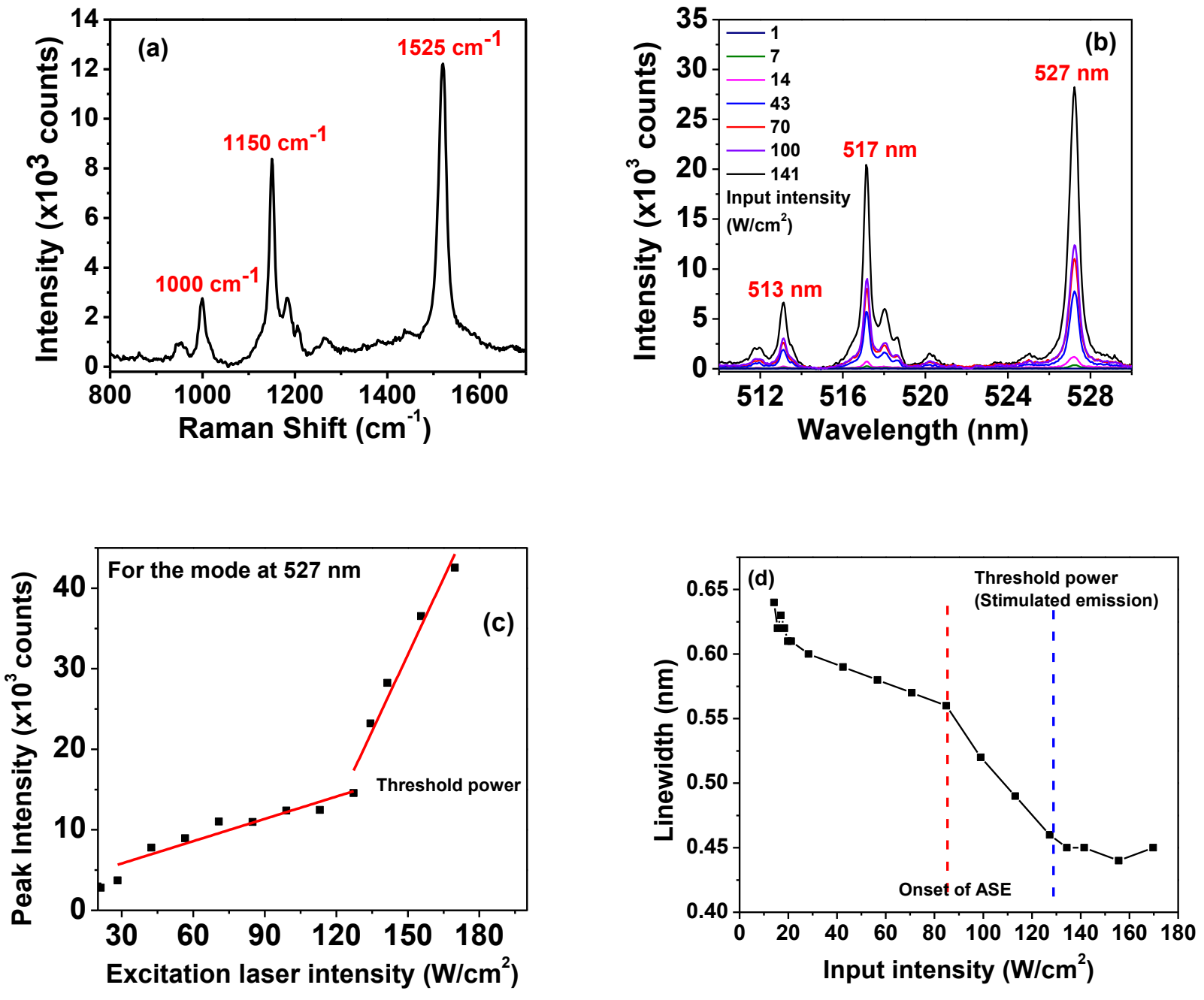

**Figure 2.** (a) Raman spectrum of carotene in carrot under 488 nm laser excitation with three stable modes, (b) intensity dependent fluorescence spectra for 488 nm excitation, (c) input-output characteristic plot corresponding to spectra in fig.2(b) showing a clear threshold power at 130 $W/cm^2$, (d) linewidth plot showing onset of ASE and stimulated emission leading to lasing.

The excitation power dependent mode intensity profile for 527 nm mode in fig. 2(c) shows a clear threshold onset at 130 $W/cm^2$ laser intensity, indicating lasing. This threshold intensity corresponds to the point where optical gain exceeds the internal non-radiative losses. Linewidth plot of intense mode at 527 nm as shown in fig.2(d) indicates a  decrease in linewidth at 90 $W/cm^2$ from 0.65 nm to 0.55 nm indicating ASE and subsequent narrowing down to 0.45 nm at the threshold power value of 130 $W/cm^2$. Although the linewidth variation is across a small range of 0.2 nm, the significant narrowing at the threshold power point provides a strong evidence for lasing characteristic at the Raman Stokes mode.

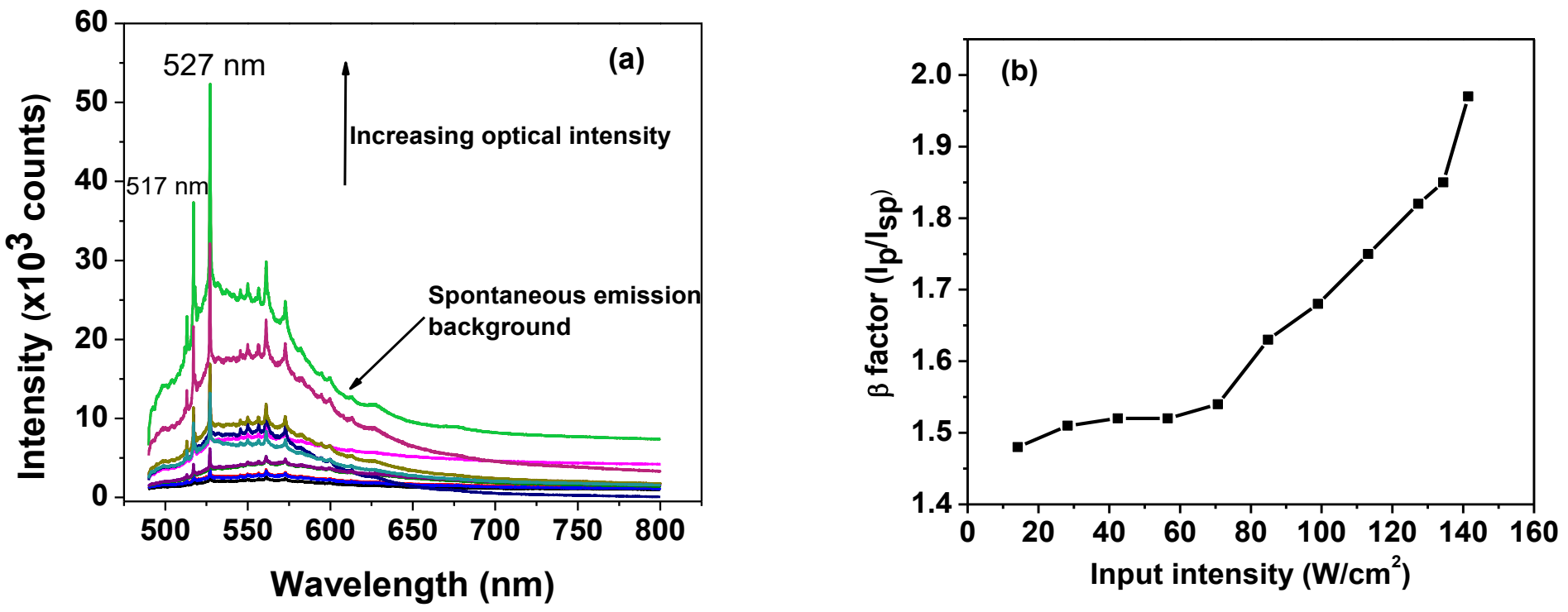

**Figure 3. (a)** Input intensity dependent emission at the 527 nm Raman mode of carotene, **(b)** $\beta$-factor plotted as a function of input pump intensity at the 527 nm lasing mode.

To probe the possible occurrence of spontaneous emission, ASE or stimulated emission in the current system, we acquired systematic power-dependent laser fluorescence spectra over a long range of 500 nm to 800 nm (fig.3(a)). The mode at 527 nm is found to exhibit considerable amplification over the spontaneous background marked. To confirm if there is any amplification at the mode, we used a parameter called $\beta$-factor[23] which is the ratio of mode intensity at 527 nm ($I_p$) to the intensity of spontaneous emission ($I_{sp}$), that is, $\beta = {I_p}/{I_{sp}}$ (shown in fig.3(b)). For normal spontaneous emission case, this ratio has to be constant with pump power. However, we see an increase in this ratio from 1.5 to almost 2 (1.9 at threshold intensity). This indicates a transition to stimulated emission at the threshold intensity. The Raman mode amplification at 527 nm may also be seeded by the PL signal at 488 nm excitation as shown in fig.1(b) earlier. There are photons between 500 nm and 600 nm peaked at around 530 nm due to fluorescence at 488 excitation. These photons help to further seed the Raman signal which is already amplified due to the Raman gain and multiple scattering. These two mechanisms, PL seeded Raman mode enhancement and stimulated emission due to multiple scattering together contribute to the observed lasing. A half-wave plate and a polarizer are used to change the input laser polarization to know the Raman signal state. The output Raman signal intensity is found to be dominant with the output polarizer in the same state as input, evidencing that the lasing mode polarization follows the pump polarization.

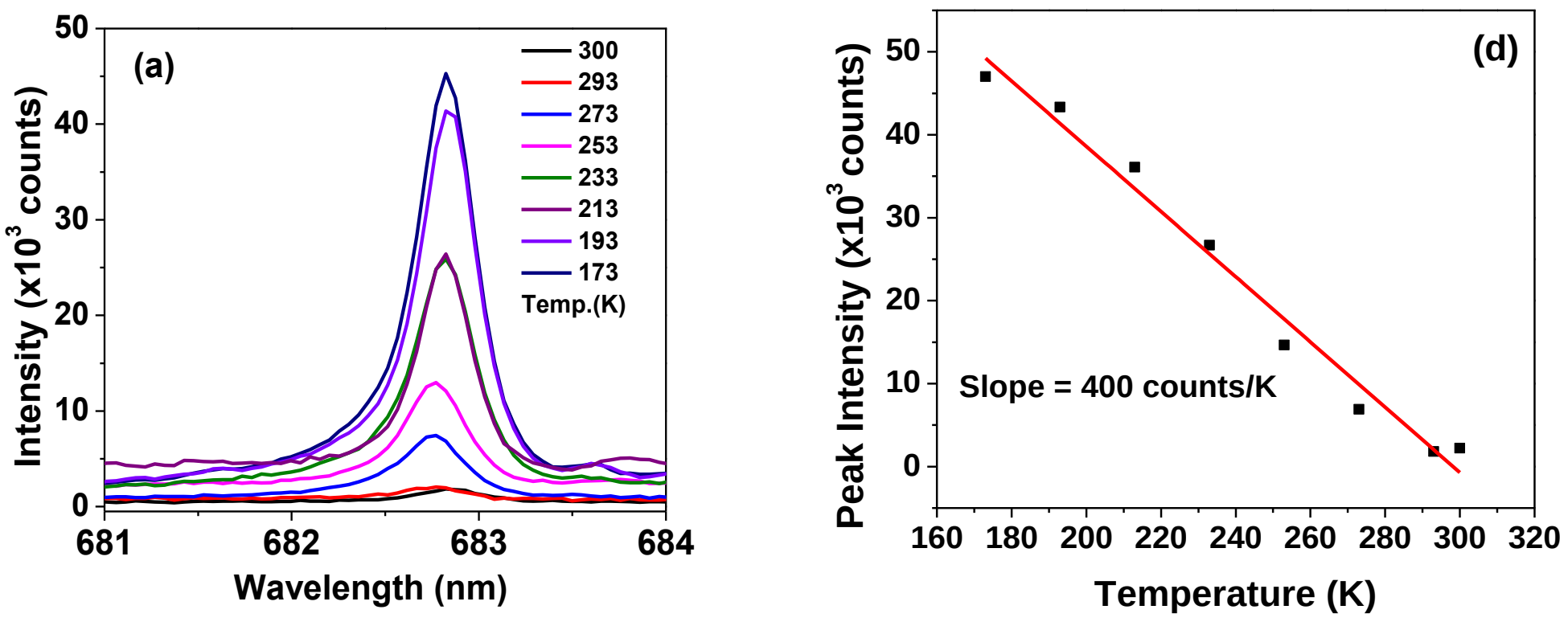

**Figure 4.** (a)Temperature variation of the peak intensities at 683 nm, (b) Linear increase in PL for mode at 700 nm is observed with decreasing temperature.

Temperature dependence on lasing mode intensity shown in fig.4(a) are obtained for temperatures varying from 173 K to 300 K ($\lambda$=632.8 nm, P=1 mW and t=1 s). PL intensity at 173 K is about 20 times of that at 300 K. The intensity variation is shown in fig.4(b), with a systematic linear trend, which could be possibly explored as a temperature sensor based on the peak mode PL intensity. Owing to enhanced Raman mode intensities due to higher phonon population and low noise at low temperatures, the emission modes can be used for sensitive applications like low-concentration analyte detection, pH and multi-parameter sensing as reported recently using random lasers[24–27]. These sensing schemes rely on mode intensity, threshold, polarization and mode wavelength variation with ambient sensing parameters which is highly feasible in our system presented here.

We have successfully demonstrated CW-laser pumped stable Stokes mode random lasing, exploiting the Raman activity of naturally occurring carotene in carrots, exhibiting a threshold of 130 W/cm$^2$ intensity, linewidth narrowing up to 0.4 nm and $Q$ factor of up to 1300. $\beta$-factor at the lasing mode and PL seeding of Raman mode signify stimulated emission. The polarization state of the lasing mode is governed by pump polarization state. The linear enhancement in PL intensity with lowering temperature from ambient conditions to 173 K points to the operational robustness of this system. These interesting findings open ways to explore further the possibility of designing such simple and bio-compatible light sources for direct biological applications involving spectroscopy, imaging and sensing.